\documentstyle[aps,psfig,prl,twocolumn]{revtex}
\begin{document}
\tightenlines
\twocolumn[\hsize\textwidth\columnwidth\hsize\csname@twocolumnfalse\endcsname
\title{Coherent Electron Transport in 
Superconducting-Normal Metallic Films}
\author{Frank K. Wilhelm$^1$, Andrei D. Zaikin$^{1,2}$ and Herv\'{e} 
Courtois$^{3}$}
\address{$^1$Institut f\"{u}r Theoretische Festk\"{o}rperphysik,
Universit\"{a}t Karlsruhe, D-76128 Karlsruhe, Germany\\
$2$ P.N. Lebedev Physics Institute, Leninskii prospect 53,
117924 Moscow, Russia\\
$3$ CRTBT-CNRS, associ\'{e} \`{a} l'Universit\'{e} Joseph Fourier, 25
Av. des Martyrs, 38042 Grenoble, France}
\maketitle
\begin{abstract}
We study the transport properties of a {\it quasi-two-dimensional} 
diffusive normal metal film attached to a superconductor. 
We demonstrate that the properties of such films can essentially 
differ from those of quasi-one-dimensional systems: in the presence
of the proximity induced superconductivity in a sufficiently wide film 
its conductance may not only increase but also {\it decrease} with temperature.
We develop a quantitative theory and discuss the physical nature of this effect.
Our theory provides a natural explanation for recent 
experimental findings referred to as the ``anomalous proximity effect''.
\end{abstract}
\pacs{73.23.Ps,74.50.+r,74.80.Fp}
]
A normal metal attached to a superconductor also acquires superconducting
properties \cite{dG}: at sufficiently low temperatures ``superconducting''
electrons penetrating into a normal metal (N) remain coherent even far from
a superconductor (S). This proximity effect can strongly influence
transport properties of the system and becomes particularly pronounced
in the case of transparent inter-metallic interfaces.

Recent theoretical and experimental studies of diffusive 
mesoscopic NS proximity structures
\cite{VZK,NazStPRL,GWZ,PetrJETPL93,PetrPRL95,Courtois,AntVolkTak}
(see also Refs. therein) revealed various interesting features of long-range
coherent states in such systems. One of such features is a non-monotonic
dependence
of the system conductance on temperature \cite{NazStPRL,GWZ,AVZ,VAL}: as the
temperature $T$ decreases below  the transition temperature $T_C$
its linear conductance $G$ increases above the normal
state value $G_N$, reaches its maximum at $T$ of the order of the
Thouless  energy $E_d$ of the normal metal and then decreases
down to $G=G_N$
at $T=0$. This non-monotonic behavior has been detected in recent
experiments \cite{Courtois}. 

The high temperature behavior of
$G(T)$ can be easily understood: as the temperature is lowered
superconductivity expands in the normal metal and
its conductance increases. The decrease of $G$ with temperature at
$T \lesssim E_d$ is due to the presence of a proximity induced
(pseudo)gap in the density of states of the N-metal at energies
$E \lesssim E_d$ \cite{GWZ}. It is also important to emphasize
that at any $0<T<T_C$ the conductance $G$ was found to be
{\it larger} than $G_N$ \cite{NazStPRL,GWZ}.

Surprisingly, in several experiments with proximity NS structures
\cite{PetrJETPL93,PetrPRL95,AntVolkTak} a {\it decrease} of the conductance
{\it below} its normal state value already {\it at the onset} of
superconductivity was observed. In some cases \cite{PetrJETPL93}
a negative correction to $G$ was as large as 30 \% of $G_N$. Even
more puzzling was the sample dependence of this result: in
\cite{PetrPRL95} a {\it decrease} of $G(T)$ with temperature
was reported if $Sb$ was chosen as a normal conductor, whereas if $Sb$ was
substituted by $Ag$ the conductance {\it increased} with
decreasing $T$. 

It appears that the explanation of the above effects based on the
assumption of low transparent NS boundaries should be ruled out:
in \cite{PetrJETPL93,PetrPRL95} 
the current does not flow directly through NS interfaces and,
on top of that, the NS boundaries in these experiments
were believed to be highly transparent.
One can also recall that in the presence of proximity
induced correlations the electric field penetrating into the normal
metal can ``overshoot'' its normal state value \cite{GWZ}. This effect,
although in principle could be interpreted as a suppression of the
local conductivity inside a part of the N-metal,
can hardly explain the results \cite{PetrJETPL93,PetrPRL95}: at
sufficiently high $T$ the ``overshooting'' effect is weak \cite{GWZ}
and is unlikely to be detected in the experimental setup
\cite{PetrJETPL93,PetrPRL95}. 
Thus it was not completely clear whether the above observations
are consistent with the existing theory of the proximity effect.

In this Letter we will develop a theory of coherent charge transport
in two-dimensional (2D) proximity metallic films. We will
demonstrate that  kinetic properties of such systems can
substantially differ from those of quasi-1D proximity structures 
\cite{NazStPRL,GWZ} due to nonuniform distribution of the current
in the film. We will show that this effect might cause a substantial
{\it decrease} of the system conductance in four-point measurements
\cite{PetrJETPL93,PetrPRL95} where the width of the samples was
of the same order as their length. We will provide a transparent
physical interpretation of the effect within a standard picture
of the proximity effect for quasi-1D normal conductors
combined with the Kirchhoff's laws. We will also discuss possible
new experiments with proximity metallic films.
\begin{figure}
\centerline{\psfig{figure=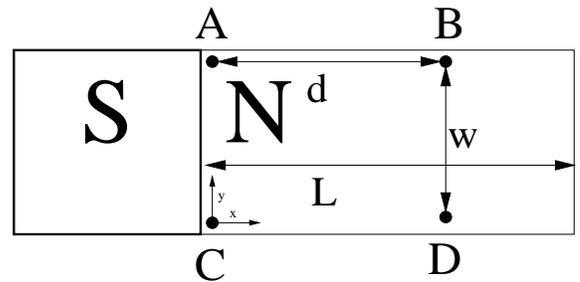,width=75mm}}
\caption{A quasi-2D proximity film. The contacts A and B are used
as voltage and C and D as current probes. An alternative setup:
A and C are voltage probes, while the current flows through B and D.}
\label{nsfilm}
\end{figure}

{\it The model and the formalism.}
Consider a planar diffusive NS-system with four probes directly
attached to the normal metal (Fig. \ref{nsfilm}). In what follows we
will assume that the NS interface as well as contacts between
probes and the N-metal are perfectly transparent. We will also
assume that the contact area between the probes and the N-metal is
small  and neglect the influence of the probes on the proximity
effect. Below we will mainly consider the following experimental
arrangement: the voltage $V$ is applied to the probes A and B, and
the current $I$ flowing in the probes C and D is measured.
A systematic description of proximity-induced coherent phenomena
in mesoscopic diffusive NS metallic structures
was obtained in \cite{VZK,NazStPRL,GWZ} within the quasiclassical
Green functions formalism of nonequilibrium superconductivity theory (see
e.g. \cite{LO}). The proximity effect can be described in a standard 
way by means of the Usadel equation \cite{Usadel}.
In the absence of inelastic scattering and interaction in the N-metal
it reads ${{\cal D}
\partial _x^2\alpha_E =-2iE\sinh \alpha_E(x)}$, where
$G_E^R=\cosh \alpha_E(x)$ and $F_E^R=\sinh \alpha_E(x)$ are the retarded normal
and anomalous Green functions and ${\cal D}$ is the diffusion
coefficient for the N-metal. In the geometry of Fig. 1 these functions
depend only on one coordinate $x$ normal to the NS interface. For
$E\ll E_{L}={\cal D}/L^2\ll\Delta$ and assuming that no current
is flowing across the metal interface at $x=L$ one readily finds
\begin{equation}
\alpha_E(x)={E\over E_{L}}{x\over L}\left(2-\frac{x}{L}\right)-i\pi/2.
\end{equation}
For $E\gg E_{L}$ superconducting correlations decay exponentially
in the normal metal and we have \cite{VZK,GWZ}
\begin{equation}
\tanh(\alpha_E/4)=\tanh(\alpha_s/4)\exp(-\sqrt{-2iE/{\cal D}}x),
\end{equation}
\begin{equation}
\alpha_s={1\over2}\tanh\left|\frac{\Delta+E}{\Delta-E}\right|-i{\pi\over2}\theta(
\Delta-E)
\end{equation}
In the absence of a supercurrent in the system the total current can
be defined as
\begin{equation}
j=\int dE\;M_E(r)\nabla f_t(r),
\end{equation}
where $f_t$ is the transverse component of the
distribution function describing deviation from equilibrium.
It satisfies the diffusion-type kinetic equation
\begin{equation}
\nabla(M_E(r)\nabla f_t)=I_E(\delta(r-r_{C})-\delta(r-r_{D})),
\label{kinetic}
\end{equation}
where $I_E$ is the spectral component of the current $I$
at the energy $E$. 
The voltage probes A and B are assumed to be in thermal
equilibrium. Then we get \cite{VZK}: $f_{tA}=0$ and
\begin{equation}
f_{tB}=\left(\tanh\left(\frac{E+V}{2T}\right)
-\tanh\left(\frac{E-V}{2T}\right)\right).
\label{reservoir}
\end{equation}
A ``no current flow'' condition at the N-metal edges yields
\begin{equation}
\partial_nf_t=0.
\label{EdgeCondition}
\end{equation}
The problem (\ref{kinetic}) is analogous to that of finding
the potential distribution in a classical
inhomogenous conductor with a local (spectral) conductivity $M_E(r)$.
Here this quantity is fully determined by the proximity effect
\begin{equation}
M_E=\sigma_N \cosh^2(Re \alpha_E(x)).
\label{loccond}
\end{equation}
where $\sigma_N$ is the normal-state conductivity.
It is important to emphasize that although the physical
picture of the proximity effect in our system is effectively
one-dimensional (and thus $M_E$ depends only on $x$), the kinetic
problem (\ref{kinetic}) is essentially {\it two-dimensional}. This
is the main difference of our model as compared to that studied in
\cite{VZK,NazStPRL,GWZ}. We will demonstrate that this feature
is crucially important leading to new physical effects.

{\it Conductance}.
A formal solution of Eq. (\ref{kinetic}) reads
\begin{equation}
f_t(E,r)=I_E({\cal G}_E(r,r_C)-{\cal G}_E(r,r_D)),
\label{formal}
\end{equation}
where ${\cal G}_E=(\nabla M_E(r)\nabla + M_E(r)\nabla ^2)^{-1}$ is
the Green function of the operator (\ref{kinetic}). Making use of
(\ref{reservoir}, \ref{EdgeCondition}) and (\ref{formal}), and
integrating $I_E$ over energies we obtain the total current $I$ and
arrive at the expression for the differential four-point-conductance $G=dI/dV$:
\begin{equation}
G(V,T)=\int_0^\infty \frac{g(E)}{2T\cosh^2((E-V)/2T)}dE ,
\end{equation}
where
\begin{equation}
g(E)=G_N\frac{{\cal G}_0^{BC}-{\cal G}_0^{BD}-{\cal
G}_0^{AC}+{\cal G}_0^{AD}}
{{\cal G}_E^{BC}-{\cal G}_E^{BD}-{\cal G}_E^{AC}+{\cal G}_E^{AD}}
\label{g}
\end{equation}
is the spectral conductance.
We introduced the notation ${\cal G}^{ij}={\cal G}(r_i,r_j)$ and
${\cal G}_0$ is the Green's function of
(\ref{kinetic}) in the normal state ($M_E(r)=\sigma_N$).
The spectral conductance (\ref{g}) was calculated numerically
from eqs. (\ref{kinetic}), (\ref{EdgeCondition}) and (\ref{loccond}).
The results are presented in Fig. 2.

For narrow films the well known results of quasi-1D calculations
\cite{NazStPRL,GWZ} are qualitatively reproduced: the linear conductance
$G(T)$ exceeds $G_N$
at all $T$ showing a non-monotonic feature at $T \lesssim E_{d}$
(for simplicity we put $L=d$ here and below).
The only quantitative
difference with \cite{NazStPRL,GWZ} occurs at low energies due to
different boundary conditions at $x=d$:
here no contact with a big normal reservoir is 
assumed and the maximum conductance $G_{max} \approx 1.12G_N$ 
is reached at $T \sim E_d/4$, i.e.
at roughly by a factor 20 lower temperature than $G_{max} \approx 1.09G_N$.
In \cite{NazStPRL,GWZ}, the proximity induced superconductivity
was slightly weaker due to the contact with a normal reservoir at $x=d$.

\begin{figure}
\centerline{\psfig{figure=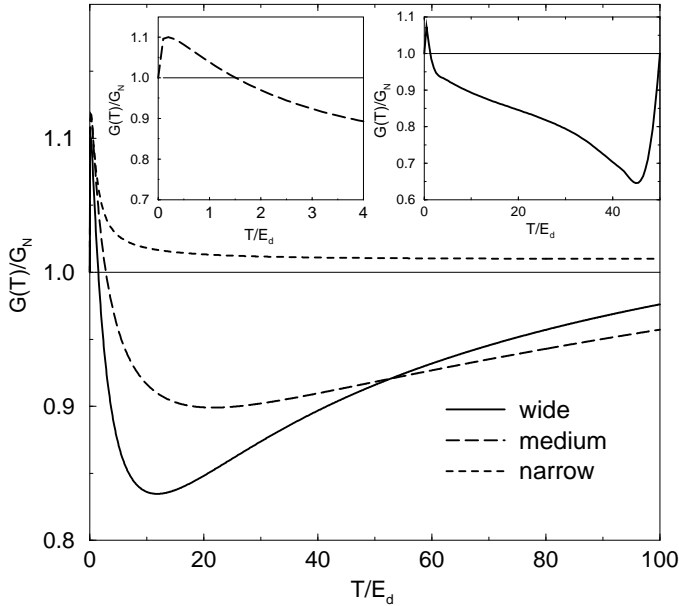,width=95mm}}
\caption{The linear conductance $G(T)$ for films of different widths
$w/d=0.05,0.5,1.0$ calculated for $d=L$ and $T_C=5.7\,10^5E_d$.
Left inset: The same curve at $w=0.5d$. The $T$-axis is zoomed to
demonstrate the presence of a usual 1D-type non-monotonic behavior at $T
\sim E_d$.
Right inset: $G(T)$ for a wide film  and $T_C=50E_d$. The amplitude
of the negative conductance peak is increased due to the effect of
a superconducting gap $\Delta (T)$.}
\label{trans2d}
\end{figure}

For broader films $G(T)$ {\it decreases} below the normal state
value at high temperatures and reaches the minimum at $T \sim
10 E_d$. At lower $T$ the conductance grows with decreasing $T$,
becomes larger than $G_N$ and then decreases again down to
$G(T=0)=G_N$ similarly to the 1D case (see the left inset in Fig.
\ref{trans2d}). The behavior of $g(E)\equiv G(E,T=0)$ as a function
of energy (voltage) is qualitatively identical to that of $G(T)$, the
negative peak at $E \sim 10E_d$ turns out to be even somewhat
deeper.
Thus we conclude that although at $T \lesssim E_{d}$ the behavior
of 2D samples essentially resembles that of a 1D system, at higher
temperatures an additional structure with the negative conductance
peak is present in the 2D case. For sufficiently wide films
the amplitude of this peak can exceed that of the positive peak
at lower $T$. This effect becomes even more pronounced if $E_d$
is not too small as compared to $T_C$ and
the peak of the density of states around the superconducting gap
should be taken into account. For typical parameters (see e.g. the
right inset in Fig. \ref{trans2d}) the minimum conductance
can be by more than 35\% smaller than $G_N$.

\begin{figure}
\centerline{\psfig{figure=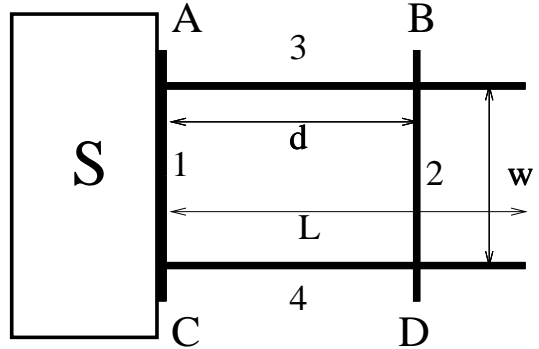,width=70mm}}
\caption{An equivalent circuit with the probe configuration as in Fig.
\ref{nsfilm}.}
\label{network}
\end{figure}

{\it The network model and current flow.}
In order to provide a transparent physical interpretation
of the above effect let us consider a simplified model of our
system: the network of quasi-1D diffusive normal wires is attached
to a superconductor as well as to current and voltage probes, see
Fig. \ref{network}. Similar equivalent cirquit model was previously
used for qualitative description of inhomogenous superconducting
films \cite{Vaglio}.
Exploiting the analogy between $f_T$ and the electrical potential in
a conventional circuit, Kirchhoff's laws for the spectral conductances
can be derived \cite{GWZ,ZaiNet}.
For the present circuit, we find (c.f. \cite{Vaglio})
\begin{equation}
g_{Net}=g_3g_4\sum_{i=1}^4 g_i^{-1}
\label{KirNet}
\end{equation}
where the $g_i$ are the spectral conductances\cite{VZK,GWZ} of the
wires 1--4
\begin{equation}
g_i=\left(\int_{\hbox{\scriptsize wire i}}{ds\over M(s)}\right)^{-1}.
\end{equation}
At $T \gg E_{d}$ only the wire 1 directly attached to a
superconductor acquires superconducting properties, whereas the
proximity effect in the wires 2, 3 and 4 is suppressed. Thus only
$g_1$ increases, and $g_{2,3,4}$ remain unaffected. According to eq.
(\ref{KirNet})  $g_{Net}$ decreases below $G_N$. At $T \lesssim
E_{d}$ the proximity induced superconducting correlation
penetrates into all four wires, $g_{2,3,4}$ increase leading to the
increase of $g_{Net}$ above $G_N$.

\begin{figure}
\centerline{\psfig{figure=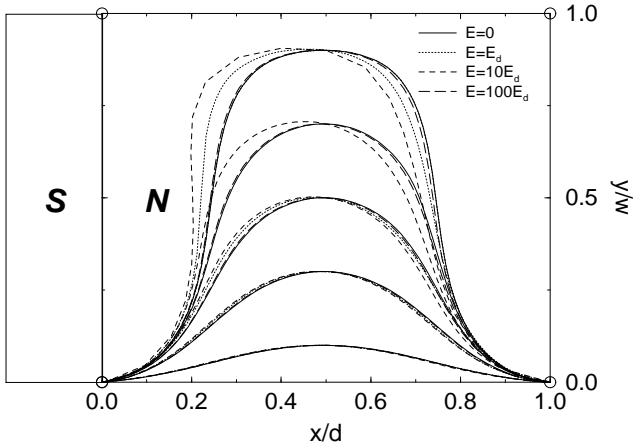,width=80mm}}
\caption{Spectral current lines in a 2D proximity film for various energies.}
\label{current}
\end{figure}

These simple arguments also suggest that the distribution of the current in
our 2D proximity system should depend on $T$: more current will flow
through ``more conducting'' parts of the N-metal.
And indeed our numerical analysis clearly demonstrates this redistribution
effect in 2D proximity films (see Fig. \ref{current}).

At low energies (where $M_E \simeq\sigma_N$) the current lines are symmetric
because the effective (spectral) conductivity $M_E \simeq\sigma_N$ is
the same everywhere in the system. At higher energies $E > E_d$
more current is flowing near the superconductor, where $M_E$ is
larger due to the proximity effect. This distorsion of the current
lines is clearly seen in Fig. \ref{current}.
At very high energies $M_E$ is increased only in
a very narrow region near the superconductor, and most current
lines become symmetric again. This illustrates the importance of the
geometry in the measuring process.

Let us finally point out that with the aid of
the above network model and the results \cite{VZK,GWZ}
one can estimate the energy $E_{cr}$, at which the crossover between
the quasi-1D ($g>G_N$) and the quasi-2D ($g<G_N$)
regimes occurs. We find that $E_{cr}\approx {{\cal D}/ w^2}$
for narrow and $E_{cr}\approx {{\cal D}/d^2}$
for wide films.
This estimate is in a good agreement with our numerical results
for 2D films.
{\it Discussion}. Our analysis clearly demonstrates that both
the temperature dependence of $G$ and the amplitude of the
effect \cite{PetrJETPL93,PetrPRL95} can be explained within
the standard quasiclassical theory of superconductivity applied to
2D proximity metallic films. This is consistent
with the fact, that in other experiments, where contacts were placed
in line \cite{Courtois}, {\em no} resistance increase below 
$T_C$ was observed. Furthermore, it also allows to
understand the sample dependence of the conductance of NS
structures observed in \cite{PetrPRL95}. 

Indeed, for the parameters of this experiment one has $E_{cr} 
\approx 10E_d \approx 40 \mu$V
and  $V\simeq R_NI \approx 7 \mu$V and 100 $\mu$V respectively for $Ag$ 
and $Sb$ samples. Thus for the $Ag$ sample $V<E_{cr}$, 
the effective 1D picture applies and the conductance {\it
increases} due to the proximity effect. On the contrary, for the
$Sb$ sample $V>E_{cr}$ and the conductance {\it decreases}
due to 2D effects. This is exactly what has been found in
\cite{PetrPRL95}. We believe that at very low voltages and
temperatures it should be possible to observe the excess
conductance effect also for $Sb$ samples.

Finally let us briefly discuss another possible 
four-point conductance measurement with different 
arrangement of voltage (A and C) and
current (B and D) probes (Fig. \ref{nsfilm}). In this case the 
spectral properties, i.e. the spread of correlations into
the normal metal, which determine $M_E(r)$, remain the same,
however the kinetics changes.
Again applying the Kirchhoff analysis we now find
\begin{equation}
g_{Net}=g_1g_2\sum_{i=1}^4g_i^{-1}
\label{KirNet2}
\end{equation}
If the voltage and current probes are close to each other,
the local conductivity is recovered. 2D effects are weak in this case since
$g_1\approx g_2$ at all energies and $g_{3,4}\ll g_{1,2}$ for $w\gg d$.
If, however, the current and voltage probes are sufficiently far
from each other one recovers two {\it positive} conductance peaks:
one at low $T \lesssim E_d$ and the second at higher $T$. The
position of this second {\it positive} conductance peak roughly
coincides with that of the {\it negative} peak ($T \sim 10E_d$) in
Fig. \ref{trans2d} for a different contact arrangement. The physical
reason for this second peak can be again understood within the
network model analysis (\ref{KirNet2}): at high enough energies
only $g_1$ is increased by the proximity effect.
These predictions agree with the results of our 2D numerical
analysis.

In conclusion, we studied kinetic properties of a 2D diffusive
normal metal film attached to a superconductor and
demonstrated that the proximity effect can lead to both
increase and decrease of the film conductance
depending on the type of measurement and the energies
involved. Our results are fully consistent with experimental
findings \cite{PetrJETPL93,PetrPRL95,AntVolkTak}. We also propose new
experiments for further study of the phenomena discussed here.

We would like to thank V.T.~Petrashov, V.N.~Antonov, B.~Pannetier, W.~Belzig,
G.~Sch\"on and A.F.~Volkov for useful discussions. This work was
supported by the Deutsche Forschungsgemeinschaft through SFB 195, 
Graduiertenkolleg ``Kollektive
Ph\"anomene im Festk\"orper'' and the INTAS-RFBR Grant 95-1305.

\end{document}